\documentclass[aps,prl,reprint,superscriptaddress,showpacs]{revtex4-2}

\usepackage{graphicx,amsmath,amssymb,subfig,mathtools,dcolumn}
\usepackage{bm}
\usepackage{upgreek}
\usepackage{float}
\begin{document}

\preprint{APS/123-QED}

\title{Enhanced Near-Field Thermal Radiation Driven by Multiple Corner and Edge Modes in Subwavelength Square Nanowires}

\author{Jose Ordonez-Miranda}
\email{jose.ordonez@cnrs.fr}
\affiliation
{Sorbonne Université, CNRS, Institut des Nanosciences de Paris, INSP, F-75005 Paris, France}

\author{Minggang Luo}
\affiliation
{Institute of Industrial Science, The University of Tokyo, Tokyo 153-8505, Japan}

\author{Michele Diego}
\affiliation
{Institute of Industrial Science, The University of Tokyo, Tokyo 153-8505, Japan}

\author{Roman Anufriev}
\affiliation
{LIMMS, CNRS-IIS IRL 2820, The University of Tokyo, Tokyo 153-8505, Japan}
\affiliation
{Institute of Industrial Science, The University of Tokyo, Tokyo 153-8505, Japan}

\author{Victor Guillemot}
\affiliation{Institute of Industrial Science, The University of Tokyo, Tokyo 153-8505, Japan}
\affiliation
{LIMMS, CNRS-IIS IRL 2820, The University of Tokyo, Tokyo 153-8505, Japan}

\author{Masahiro Nomura}
\affiliation{Institute of Industrial Science, The University of Tokyo, Tokyo 153-8505, Japan}
\affiliation
{LIMMS, CNRS-IIS IRL 2820, The University of Tokyo, Tokyo 153-8505, Japan}

\author{Sebastian Volz}
\affiliation
{Laboratoire des Solides Irradiés Ecole Polytechnique - UMR CNRS 7642 - CEA/DRF/IRAMIS 28 route de Saclay, 91128 Palaiseau Cedex, France}

\date{\today}

\begin{abstract}
We demonstrate that the near-field thermal radiation between subwavelength SiC nanowires with square cross sections is dominated by multiple corner and edge resonances rather than the single surface-phonon-polariton channel of planar surfaces. Fluctuational electrodynamics simulations reveal that these resonances lie within the SiC Reststrahlen band, redshift for thinner nanowires, and yield a four-fold enhancement of thermal conductance. This maximum enhancement occurs when the separation gap nearly matches the nanowire thickness, balancing dimensional confinement and interwire coupling. These findings establish square nanowires as a versatile platform for geometry-controlled near-field heat transfer in nanoscale heat management and energy conversion.
\end{abstract}
\maketitle
 
The manipulation of near-field radiative heat transfer (NFRHT) has emerged as a transformative approach for advancing thermal management, energy conversion, and nanoscale photonic technologies. Over the past decade, many research groups have demonstrated that NFRHT between planar surfaces and subwavelength membranes can exceed the blackbody limit by orders of magnitude, primarily due to the tunneling of evanescent electromagnetic waves mediated by surface phonon-polaritons (SPhPs) and localized electromagnetic modes \cite{Polder1971,Joulain2005,Rousseau2009,Song2016,DeSutter2019,Shi2019,Shi2021,Wenbin2024}. The recent exploration of NFRHT between nanostructures with complex geometries has revealed even richer physics, including the emergence of corner and edge modes that drive heat transfer when two dimensions of the radiating system are much smaller than the dominant wavelength of thermal radiation ($\lambda_{th}\approx 10~\upmu$m at 300~K) \cite{Luo2024,Tang2024,Zeng2025,Correa2026}. 

Recent studies have shown that corner and edge modes, sustained by subwavelength membranes and cylinders, can dramatically enhance the NFRHT compared to infinite planar surfaces. For example, Tang et al. \cite{Tang2024} observed a substantial enhancement in the radiative heat transfer coefficient between 20-nm-thick silicon carbide (SiC) membranes, attributed to the coupling of two corner and edge modes. Similarly, Zeng et al. \cite{Zeng2025} reported that edge modes in subwavelength cylinders with circular cross sections can dominate the NFRHT and significantly surpass the blackbody limit. 
More recently, McCormack et al. \cite{Correa2026} showed that two corner and edge modes enhance the NFRHT between low-loss polar membranes like SiC and attenuate it for high-loss ones, such as SiO$_2$, due to the material energy absorption reducing the density of electromagnetic states. The nanostructures considered in all these studies \cite{Tang2024,Zeng2025,Correa2026} sustain a limited number of corner and edge modes, whose radiative contribution peaks in the dual nanoscale regime, where both thickness (diameter) and separation gap of the membranes (cylinders) are much smaller than $\lambda_{th}$. These previous results highlight the great potential of dimensional confinement to tailor nanoscale thermal radiation via corner and edge modes. Therefore, the transformation of nanofilms into square-cross-section nanowires via the reduction of both their thickness and width is expected to naturally sustain enhanced coupling of highly localized electromagnetic modes at corners and edges. However, the NFRHT physics in square nanowires separated by subwavelength gaps (triple nanoscale regime) remains unexplored, despite its promising suitability for supporting rich multimode interactions and strong thermal radiation.

Here, we demonstrate that subwavelength SiC nanowires with square cross sections support multiple corner and edge resonances that dominate the NFRHT over SPhPs. Using fluctuational electrodynamics simulations, we find that the evanescent coupling of these electromagnetic modes maximizes the NFRHT when the separation gap of the nanowires nearly matches their thickness. Thinner nanowires sustain more resonances with redshifted frequencies and higher intensities, which highlights the crucial role of the triple dimensional confinement (thickness, width, gap) to tune the radiative spectrum and enhance nanoscale thermal radiation. 

We consider two nanowires exchanging heat via thermal radiation due to their temperature difference $\Delta T$, as shown in Fig.~1. The nanowires have square cross sections to naturally promote the emergence of corner and edge modes, while their subwavelength thickness $t=10-300~$nm was chosen small enough to ensure the cross-plane coupling and in-plane propagation of SPhPs. The nanometric separation distance $d= 10-300~$nm was varied one order of magnitude to identify optimal conditions for enhancing the thermal radiation. These subwavelength values of both $t$ and $d$ are expected to strengthen the impact of the corner and edge modes \cite{Tang2024}. In addition, the nanowires' length $l=500~$nm was selected sufficiently long to sustain the in-plane propagation of surface electromagnetic waves and small enough to keep the numerical calculations tractable. The study was done for nanowires of SiC, a polar material able to support the excitation and propagation of SPhPs as well as corner and edge modes \cite{Volz24,Ordonez2026,Tang2024}. These nanowires with a square cross section can be fabricated by the top-down \cite{Anufriev2022} and bottom-up \cite{Chauvin2012} approaches.

\begin{figure}
\centering
\includegraphics[width=1\linewidth]{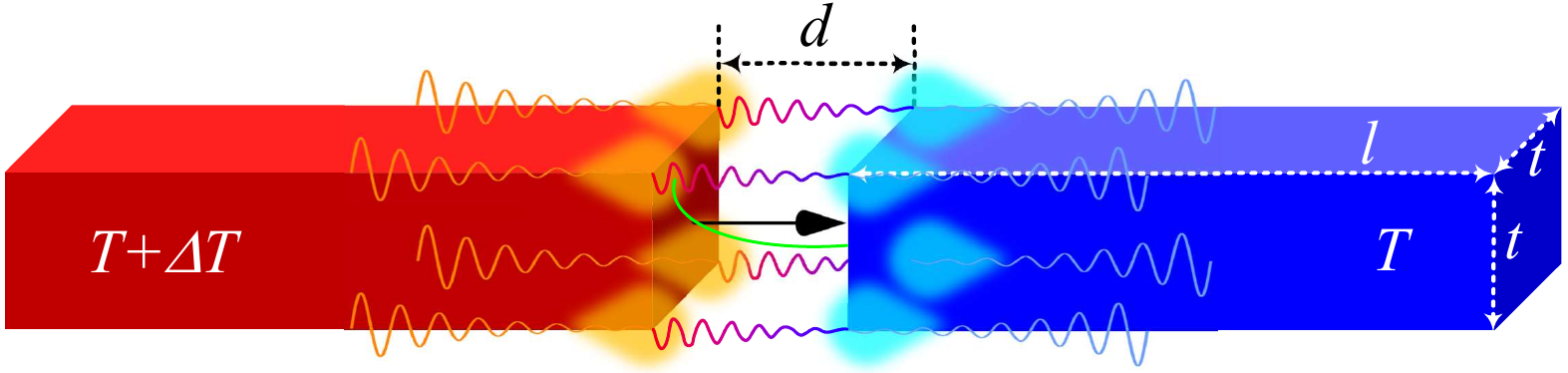}
\caption{Scheme of two subwavelength nanowires exchanging thermal radiation driven by propagating (black arrow), evanescent (Green line), and corner/edge (yellow and light blue clouds) electromagnetic modes. These latter modes result from the interference between the incident and reflected SPhPs (orange and light blue waves) at the nanowire boundaries supporting the evanescent transmission (reddish wavy lines) of their energy through the gap. Both the thickness and gap of the nanowires are smaller than the dominant thermal wavelength ($t,d\ll \lambda_{th}$) to ensure the appearance of corner and edge modes. The red and blue colors of the identical nanowires indicate high ($T+\Delta T$) and low ($T$) temperatures, respectively. }
\label{fig1}
\end{figure}

The uniform temperature of each nanowire generates the internal fluctuation of electromagnetic currents, whose energy emission results in thermal radiation \cite{Joulain2005,Volz24}. Considering that the temperature difference between the nanowires is relatively small ($\Delta T\ll T$), the net heat rate $q=G\Delta T$ between the nanowires is determined by the thermal conductance $G$ given by fluctuational electrodynamics, as follows \cite{Basu09,Ordonez24}
\begin{equation}
	G=\int_{0}^{\infty}\hbar\omega \frac{\partial f_{\omega}(T)}{\partial T}\Phi(\omega)d\omega.
\end{equation}
where $\omega$ is the radiation frequency, $f_{\omega}(T)=\left[\exp\left(\hbar\omega/k_B T\right)-1 \right]^{-1}$ is the Bose-Einstein distribution function, $\hbar$ and $k_B$ are the respective reduced Planck and Boltzmann constants, and $\Phi$ is the transmission function of the energy emitted from the hot nanowire (emitter) into the cold one (receiver). According to the Landauer formalism \cite{Biehs21}, $\Phi$ accounts for the sum of the transmission probabilities of each radiative channel (i. e. Propagating, evanescent, corner, and edge modes with TM and TE polarizations) and therefore its values can be greater than unity \cite{Ordonez2026}. The frequency spectrum of $\Phi(\omega)$ is determined by the electromagnetic fields generated by the fluctuation of surface and volumetric currents, which we calculate through the boundary-element method \cite{Rodriguez13}. The predictions of this method have been validated by abundant experimental data obtained for both the near-field and far-field regimes of thermal radiation \cite{Thompson18,Luo2024,Tachikawa24,Ordonez24,Ordonez25}, and we used it here via the open-source software Scuff-EM. The calculations were done by using the complex SiC permittivity ($\varepsilon$) and discretizing both nanowires into surface elements, as shown in Figs.~S2 and S4(a) of the supplementary material (SM), respectively. The Reststrahlen band ($\text{Re}(\varepsilon)<0$) of SiC spans from its transversal optical frequency ($\omega_T/2\pi=23.71~$THz) to its longitudinal one ($\omega_L/2\pi=28.97~$THz), which determine the peak and dip of the amplitude of $\varepsilon$, respectively. The mesh size ($4.2~$nm) was chosen to be much smaller than the relevant radiative wavelengths ($>1~\upmu$m at 300~K) and dimensions of the radiating system to ensure the simulations' convergence, as detailed in the supplementary Fig.~S4(b).
\begin{figure}
	\centering
	\includegraphics[width=1\linewidth]{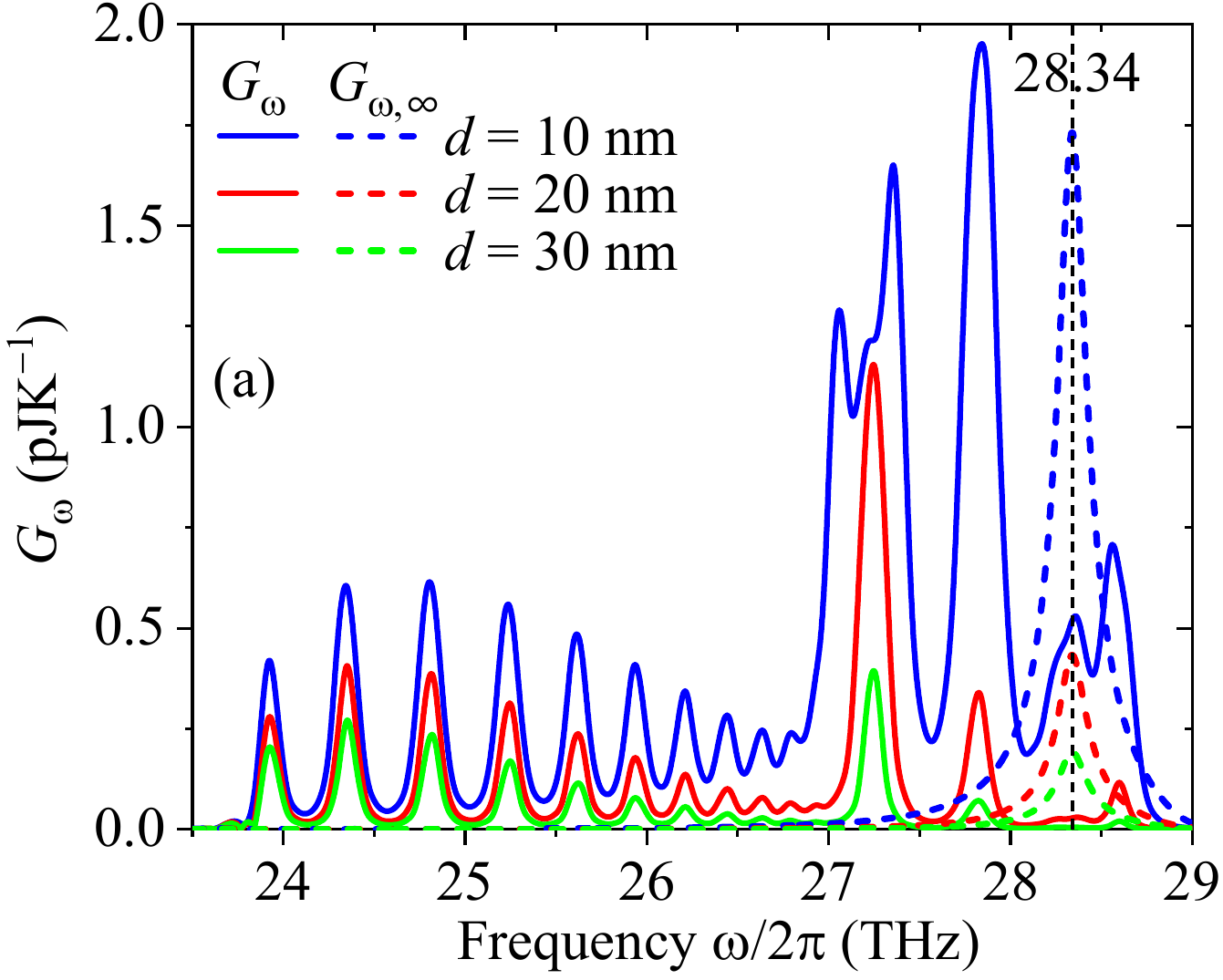}
	\includegraphics[width=1\linewidth]{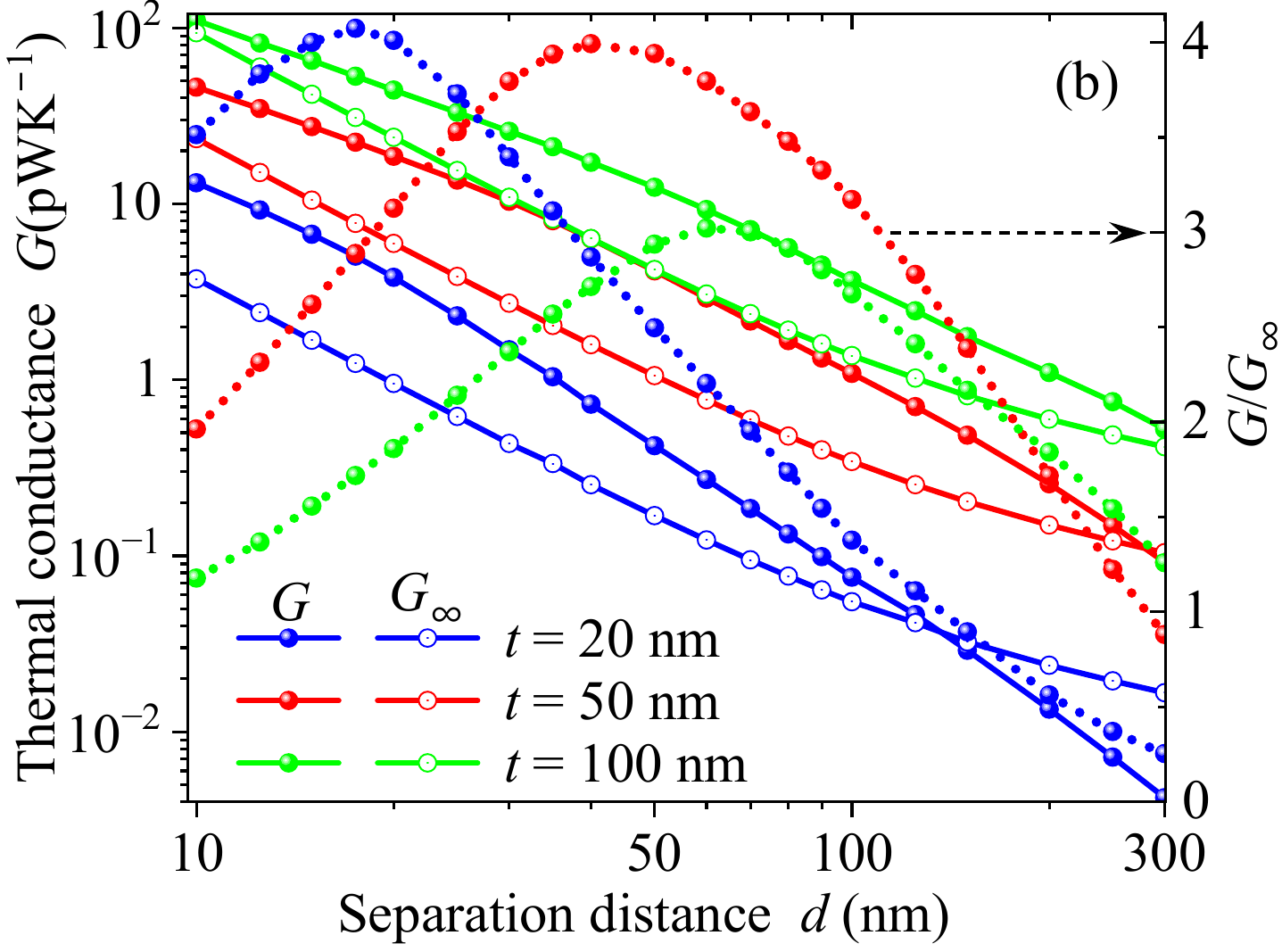}
	\caption{(a) Thermal conductance spectra $G_{\omega}$  and  $G_{\omega,\infty}$ along with (b) their integrated counterparts $G$ and $G_{\infty}$ as functions of the separation distance of two nanowires and infinite surfaces of SiC, respectively. The dotted lines in (b) represent the ratio $G/G_{\infty}$. Calculations were done for room temperature ($T=300~$K) and three representative separation distances and thicknesses. The nanowire thickness $t=20~$nm was used in (a).}
	\label{fig2}
\end{figure}

The spectral thermal conductances for two 20-nm-thick nanowires ($G_{\omega}$) and two parallel elements of infinite planar surfaces ($G_{\omega,\infty}$) are comparatively shown in Fig.~2(a), for three near-field separation distances: $d=10, 20, 30~$nm. For comparison, we chose the cross-section area ($t^2$) of these elements to be equal to that of the nanowires, as detailed in the SM section 3. The frequency spectrum of $G_{\omega,\infty}$ exhibits a single and broad peak centered at $\omega/2\pi=28.34~$THz, which falls within the SiC Reststrahlen band and determines the condition $\text{Re}(\varepsilon)=-1$. This resonance frequency maximizes the density of states and minimizes the in-plane and cross-plane propagation lengths of SPhPs at a SiC/vacuum interface, as shown in the supplementary Fig.~S3. Therefore, the characteristic peak of $G_{\omega,\infty}$ provides the main contribution to the near-field thermal radiation dominated by the coupling of SPhPs across the gap of planar surfaces. In stark contrast, the spectral conductance of nanowires reveals multiple sharp resonant peaks inside the Reststrahlen band. These low-frequency ($\omega/2\pi<28.34~$THz) peaks arise from the strong dimensional confinement of the square nanowires. For a small separation distance ($d=10~$nm), the primary peak shows up near the SPhP resonance frequency ($\omega/2\pi=28.34~$THz), whereas for larger gaps ($d=20$ and $30~$nm), it is redshifted to a lower frequency ($27.24~$THz). The cluster of secondary peaks appears for $\omega/2\pi<27.24~$THz with higher amplitudes for smaller gaps. The frequencies of these peaks are independent of $d$ and satisfy the second-order recurrence relation $\omega_{n+1}=2.31\omega_{n}-1.31\omega_{n-1}$, with $\omega_{0}/2\pi=23.93~$THz and $\omega_{1}/2\pi=24.35~$THz. Crucially, for all these small gaps ($d\leq 30~$nm), the area ($G=\int G_{\omega} d\omega$) below the primary and secondary peaks of $G_{\omega}$ clearly exceed the corresponding one of the single peak of $G_{\omega,\infty}$, which indicates that these resonances are the main mechanism of NFRHT between subwavelength nanowires. This multipeak spectrum represents a significant departure from the conventional thermal radiation driven by SPhP resonances in planar systems, highlighting the unique role of the triple geometric confinement.

\begin{figure}
	\centering
	\includegraphics[width=1\linewidth]{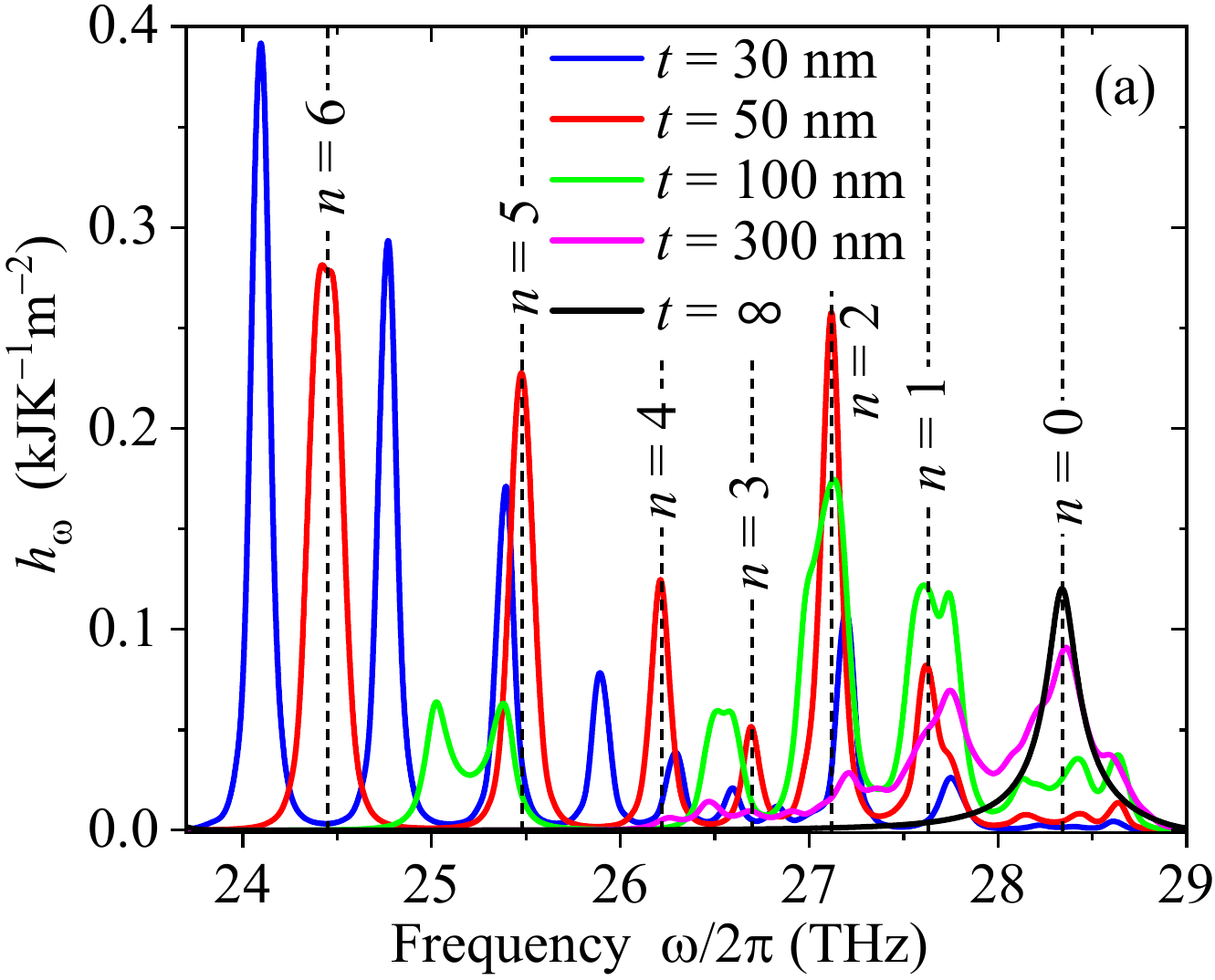}
	\includegraphics[width=1\linewidth]{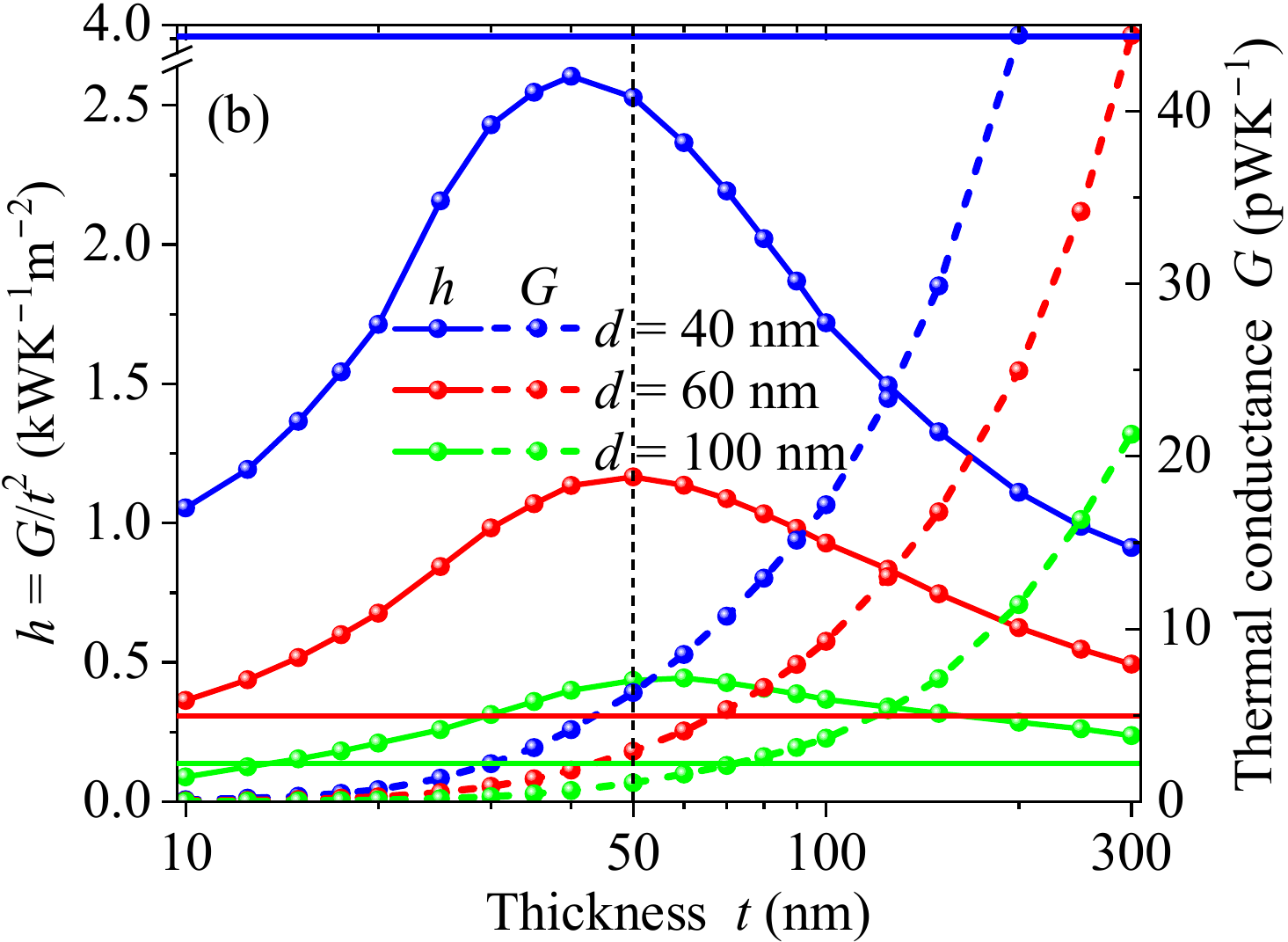}
	\caption{(a) Frequency spectrum of the coefficient of thermal radiation  $h_{\omega}$  and (b) its integrated counterpart $h$ as a function of the thickness of two SiC nanowires. The horizontal lines in (b) represent the values of $h$ for infinite SiC surfaces ($t=\infty$), serving as baselines for comparison. Calculations were done for room temperature ($T=300~$K) and representative thicknesses and separation distances. The gap thickness $d=60~$nm was used in (a), where the six peaks of $h_{\omega}$ for $t=50~$nm were labeled with $n=1,2,...6$. For reference, the single spectral peak of infinite surfaces was labeled with $n=0$.}
	\label{fig3}
\end{figure}

Figure 2(b) shows the total thermal conductances $G$ and $G_{\infty}$ as functions of the separation distance $d$ of nanowires and infinite surfaces, respectively. For the three considered thicknesses ($t=20, 50, 100~$nm), $G$ is consistently higher than $G_{\infty}$ and decays more slowly with increasing $d$. The highest conductance enhancement ($G/G_{\infty}$) over infinite surfaces, occurs at intermediate gaps, where the non-SPhP modes play a dominant role over evanescent (SPhPs) and propagating modes. The thinnest nanowires ($t=20~$nm) exhibit the highest ratio $G/G_{\infty}>4$ at $d=17.5~$nm, reflecting the pronounced impact of dimensional confinement to amplify the radiative contribution of these modes. Although $G/G_{\infty}$ decreases as $d$ increases, the nanowires maintain a superior thermal conductance even beyond $d=100~$nm, indicating the persistence of these modes over extended gaps. As the nanowires' thickness increases ($t=50, 100~$nm), the enhancement peak weakens and shifts to larger gaps comparable to and smaller than thickness ($d\lesssim t$), which reveals the close values of $t$ and $d$ that optimize the evanescent coupling of non-SPhP modes. 
\begin{figure*}
	\centering
	\includegraphics[width=1\linewidth]{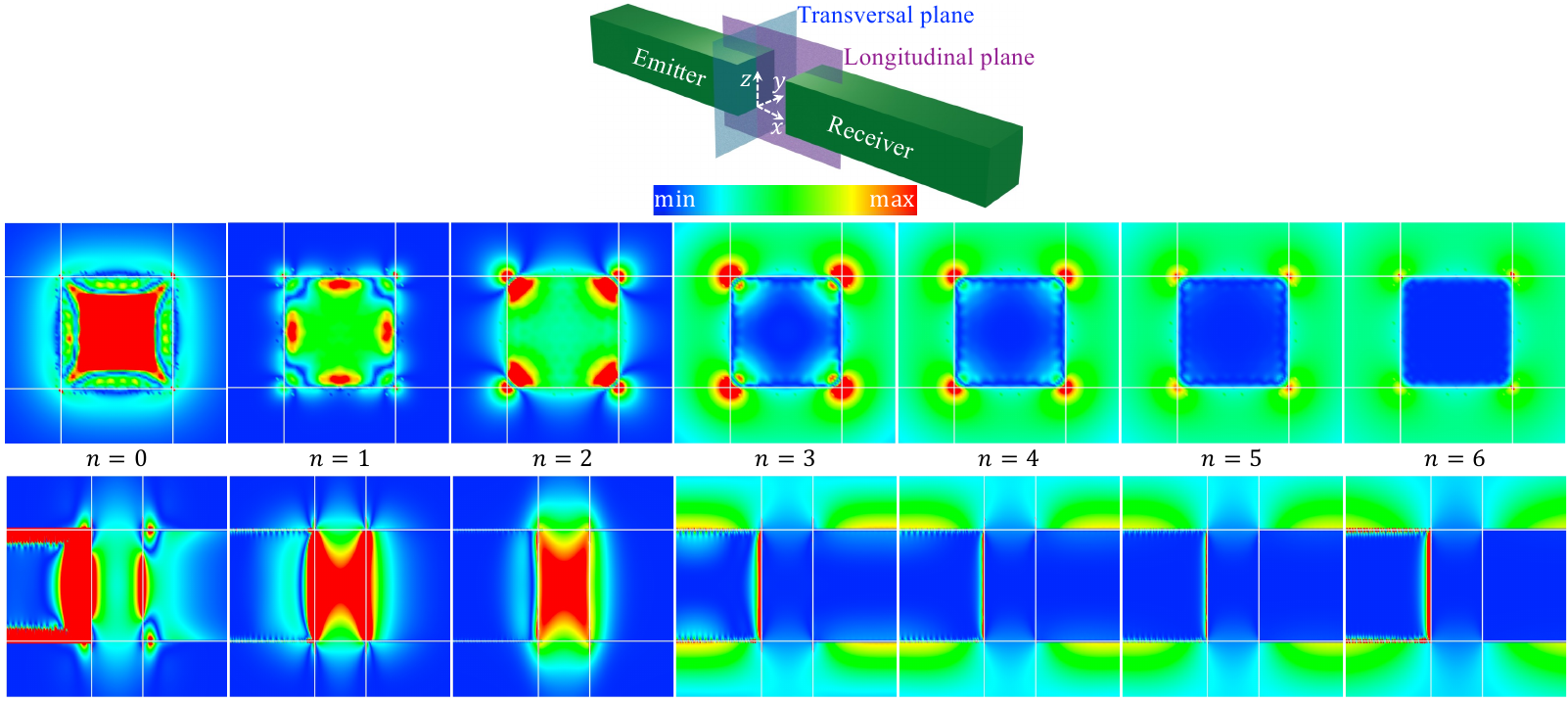}
	\caption{Density maps of the Poynting vector component $S_x$ calculated in the transversal and longitudinal planes (top) $x=0$ and (bottom) $y=0~$nm of two 50-nm-thick nanowires of SiC separated by a distance $d=60~$nm. These cuts of the energy density $S_x$ enable visualizing the localization and geometry of the corner and edge modes along with their transmission across the separation gap. Calculations were done with Scuff-EM for the peak frequencies $n=0,1,...,6$ of the spectral coefficient of thermal radiation shown in Fig.~3(a). The horizontal and vertical white lines represent the surface positions and their lengths are in the scale $x:y=1.6:1$, for the bottom panel.}
	\label{fig4}
\end{figure*}

To further investigate the origin of the $d\lesssim t$ optimum in Fig.~2(b), we examine how the heat transfer coefficient $h=G/t^2$ evolves with the nanowire thickness $t$ for a fixed gap. Figure 3(a) shows that the spectral coefficient $h_{\omega}=G_{\omega}/t^2$ of thinner nanowires supports a larger number of resonant peaks inside the Reststrahlen band. This multi-resonant behavior of nanowires is redshifted from the single-peak spectrum of planar surfaces ($t=\infty$, $n=0$), in agreement with the trend shown in Fig.~2(a) for different separation distances. For the thinnest nanowires ($t=30~$nm), the peaks are more closely spaced and exhibit higher amplitudes, indicating stronger confinement and enhanced contributions from the non-SPhP modes. As $t$ increases, the peaks broaden and shift to higher frequencies, reflecting the reduced influence of dimensional confinement. For $t = 50~$nm, the resonant frequencies satisfy the condition $\omega_{n+1}=1.58\omega_{n}-0.58\omega_{n-1}$, with $\omega_{0}=28.34~$THz and $\omega_{1}=27.63~$THz, for $n=0,1,...,6$, as shown in Fig.~3(a). The coefficients of this second-order recurrence relation are smaller than the corresponding ones found for $t=20~$nm (see Fig.~2(a)), consistent with the redshift observed for thinner nanowires. Importantly, the multi-peak spectra in Fig.~2(a) and 3(a) are a distinctive feature of the thermal radiation between nanowires, as they are absent for nanofilms of the same thickness \cite{Correa2026,Ordonez2026}. A similar cluster of peaks also appears for circular nanowires (see section 4 and 6 of SM), which reveals the key role of reducing the in-plane dimensions of nanofilms to increase the number of radiative resonances. This dimensional confinement  results in a pronounced peak of $h$, for intermediate thicknesses, as shown in Fig.~3(b). For $d=40~$nm, the peak appears at $t=40~$nm (gap $=$ thickness), while for $d=60~$nm, the peak occurs at $t=50~$nm. Larger gaps yield shorter peaks appearing for a thickness close to and smaller than the separation gap ($t\leq d$). The $h$ peak thus corresponds to the optimal balance between dimensional confinement and coupling of the non-SPhP modes driving the thermal radiation. As $t$ increases beyond its peak value, $h$ decreases and approaches the infinite surface limit, indicating a transition from a corner/edge mode-dominated regime to a SPhP-driven regime. This trend of $h$ is consistent with the one of $G$ identified in Fig.~2(b), where the highest enhancement $G/G_{\infty}$ occurs when gap nearly matches the thickness, confirming that both the longitudinal and transversal confinement jointly control the contribution of the non-SPhP modes. 

Figure 4 shows the spatial distribution of the Poynting vector component $S_x$ in the transversal ($x=0$, top panel) and longitudinal ($y=0~$, bottom panel) planes of two 50-nm-thick SiC nanowires separated by a 60-nm vacuum gap, computed at the resonant frequencies $n=0$ to 6 from Fig.~3(a). The intense energy concentration at the nanowires' surfaces, edges, and corners clearly illustrates the localization of SPhP ($n=0$), edge ($n=1$), and corner ($n=2,...,6$) modes, respectively. The non-SPhP modes thus correspond to corner and edge modes naturally sustained by the square cross section of the nanowires. The bottom-panel maps reveal the evanescent decay of these modes across the gap, yet with sufficient amplitude to enable a substantial thermal transfer between the nanowires, especially for the lower-order modes $n=1$ and $2$. These modes decay more slowly than the higher-frequency SPhP mode ($n=0$), promoting stronger interwire coupling and enhanced energy transmission across the gap. Thus, the subwavelength dimensional confinement of nanowires with a square cross section, concentrates energy at corners and edges, and enable multi-channel tunneling, boosting the NFRHT as evidenced in Figs.~2 and 3 and section 7 of the SM. This dominance of highly localized corner and edge modes marks a paradigm shift from the traditional SPhP-driven thermal radiation in planar geometries.

In summary, we have demonstrated that the near-field thermal radiation between subwavelength SiC nanowires with square cross-sections is dominated by corner and edge electromagnetic modes. Based on fluctuational electrodynamics, we have found that these localized modes generate multiple radiative resonances within the SiC Reststrahlen band, resulting in a four-fold enhancement of the thermal radiation with respect to that between planar surfaces. Thinner nanowires support a larger number of modes redshifted from the polariton peak, which highlights their geometric origin and the tunability of thermal radiation. The balance between the dimensional confinement and coupling of the corner and edge modes, generates a maximum exhancement for a thickness comparable to the separation gap. These findings provide a robust framework for designing thermal photonic devices and could be extended for nanowires of other polar materials and metals supporting the propagation of surface polaritons along with corner and edge modes. 

\vspace*{1.5cm}
\begin{acknowledgments}
This work was supported by the CREST JST (Grant N° JPMJCR19I1) and KAKENHI JSPS (Grant N° 21H04635 and JP20J13729) projects.
\end{acknowledgments}

\newpage

\begin{thebibliography}{23}%
	\makeatletter
	\providecommand \@ifxundefined [1]{%
		\@ifx{#1\undefined}
	}%
	\providecommand \@ifnum [1]{%
		\ifnum #1\expandafter \@firstoftwo
		\else \expandafter \@secondoftwo
		\fi
	}%
	\providecommand \@ifx [1]{%
		\ifx #1\expandafter \@firstoftwo
		\else \expandafter \@secondoftwo
		\fi
	}%
	\providecommand \natexlab [1]{#1}%
	\providecommand \enquote  [1]{``#1''}%
	\providecommand \bibnamefont  [1]{#1}%
	\providecommand \bibfnamefont [1]{#1}%
	\providecommand \citenamefont [1]{#1}%
	\providecommand \href@noop [0]{\@secondoftwo}%
	\providecommand \href [0]{\begingroup \@sanitize@url \@href}%
	\providecommand \@href[1]{\@@startlink{#1}\@@href}%
	\providecommand \@@href[1]{\endgroup#1\@@endlink}%
	\providecommand \@sanitize@url [0]{\catcode `\\12\catcode `\$12\catcode
		`\&12\catcode `\#12\catcode `\^12\catcode `\_12\catcode `\%12\relax}%
	\providecommand \@@startlink[1]{}%
	\providecommand \@@endlink[0]{}%
	\providecommand \url  [0]{\begingroup\@sanitize@url \@url }%
	\providecommand \@url [1]{\endgroup\@href {#1}{\urlprefix }}%
	\providecommand \urlprefix  [0]{URL }%
	\providecommand \Eprint [0]{\href }%
	\providecommand \doibase [0]{https://doi.org/}%
	\providecommand \selectlanguage [0]{\@gobble}%
	\providecommand \bibinfo  [0]{\@secondoftwo}%
	\providecommand \bibfield  [0]{\@secondoftwo}%
	\providecommand \translation [1]{[#1]}%
	\providecommand \BibitemOpen [0]{}%
	\providecommand \bibitemStop [0]{}%
	\providecommand \bibitemNoStop [0]{.\EOS\space}%
	\providecommand \EOS [0]{\spacefactor3000\relax}%
	\providecommand \BibitemShut  [1]{\csname bibitem#1\endcsname}%
	\let\auto@bib@innerbib\@empty
	\bibitem [{\citenamefont {Polder}\ and\ \citenamefont
		{Van~Hove}(1971)}]{Polder1971}%
	\BibitemOpen
	\bibfield  {author} {\bibinfo {author} {\bibfnamefont {D.}~\bibnamefont
			{Polder}}\ and\ \bibinfo {author} {\bibfnamefont {M.}~\bibnamefont
			{Van~Hove}},\ }\bibfield  {title} {\bibinfo {title} {Theory of radiative heat
			transfer between closely spaced bodies},\ }\href
	{https://doi.org/10.1103/PhysRevB.4.3303} {\bibfield  {journal} {\bibinfo
			{journal} {Phys. Rev. B}\ }\textbf {\bibinfo {volume} {4}},\ \bibinfo {pages}
		{3303} (\bibinfo {year} {1971})}\BibitemShut {NoStop}%
	\bibitem [{\citenamefont {Joulain}\ \emph {et~al.}(2005)\citenamefont
		{Joulain}, \citenamefont {Mulet}, \citenamefont {Marquier}, \citenamefont
		{Carminati},\ and\ \citenamefont {Greffet}}]{Joulain2005}%
	\BibitemOpen
	\bibfield  {author} {\bibinfo {author} {\bibfnamefont {K.}~\bibnamefont
			{Joulain}}, \bibinfo {author} {\bibfnamefont {J.-P.}\ \bibnamefont {Mulet}},
		\bibinfo {author} {\bibfnamefont {F.}~\bibnamefont {Marquier}}, \bibinfo
		{author} {\bibfnamefont {R.}~\bibnamefont {Carminati}},\ and\ \bibinfo
		{author} {\bibfnamefont {J.-J.}\ \bibnamefont {Greffet}},\ }\bibfield
	{title} {\bibinfo {title} {Surface electromagnetic waves thermally excited:
			Radiative heat transfer, coherence properties and casimir forces revisited in
			the near field},\ }\href {https://doi.org/10.1016/j.surfrep.2004.10.003}
	{\bibfield  {journal} {\bibinfo  {journal} {Surf. Sci. Rep.}\ }\textbf
		{\bibinfo {volume} {57}},\ \bibinfo {pages} {59} (\bibinfo {year}
		{2005})}\BibitemShut {NoStop}%
	\bibitem [{\citenamefont {Rousseau}\ \emph {et~al.}(2009)\citenamefont
		{Rousseau}, \citenamefont {Siria}, \citenamefont {Jourdan}, \citenamefont
		{Volz}, \citenamefont {Comin}, \citenamefont {Chevrier},\ and\ \citenamefont
		{Greffet}}]{Rousseau2009}%
	\BibitemOpen
	\bibfield  {author} {\bibinfo {author} {\bibfnamefont {E.}~\bibnamefont
			{Rousseau}}, \bibinfo {author} {\bibfnamefont {A.}~\bibnamefont {Siria}},
		\bibinfo {author} {\bibfnamefont {G.}~\bibnamefont {Jourdan}}, \bibinfo
		{author} {\bibfnamefont {S.}~\bibnamefont {Volz}}, \bibinfo {author}
		{\bibfnamefont {F.}~\bibnamefont {Comin}}, \bibinfo {author} {\bibfnamefont
			{J.}~\bibnamefont {Chevrier}},\ and\ \bibinfo {author} {\bibfnamefont
			{J.-J.}\ \bibnamefont {Greffet}},\ }\bibfield  {title} {\bibinfo {title}
		{Radiative heat transfer at the nanoscale},\ }\href
	{https://doi.org/10.1038/nphoton.2009.94} {\bibfield  {journal} {\bibinfo
			{journal} {Nat. Photonics}\ }\textbf {\bibinfo {volume} {3}},\ \bibinfo
		{pages} {514} (\bibinfo {year} {2009})}\BibitemShut {NoStop}%
	\bibitem [{\citenamefont {Song}\ \emph {et~al.}(2016)\citenamefont {Song},
		\citenamefont {Ganjyal}, \citenamefont {Sadat}, \citenamefont {Reddy},\ and\
		\citenamefont {Meyhofer}}]{Song2016}%
	\BibitemOpen
	\bibfield  {author} {\bibinfo {author} {\bibfnamefont {B.}~\bibnamefont
			{Song}}, \bibinfo {author} {\bibfnamefont {G.}~\bibnamefont {Ganjyal}},
		\bibinfo {author} {\bibfnamefont {S.}~\bibnamefont {Sadat}}, \bibinfo
		{author} {\bibfnamefont {P.}~\bibnamefont {Reddy}},\ and\ \bibinfo {author}
		{\bibfnamefont {E.}~\bibnamefont {Meyhofer}},\ }\bibfield  {title} {\bibinfo
		{title} {Radiative heat conductances between dielectric and metallic parallel
			plates with nanoscale gaps},\ }\href {https://doi.org/10.1038/nnano.2016.46}
	{\bibfield  {journal} {\bibinfo  {journal} {Nat. Nanotechnol.}\ }\textbf
		{\bibinfo {volume} {11}},\ \bibinfo {pages} {509} (\bibinfo {year}
		{2016})}\BibitemShut {NoStop}%
	\bibitem [{\citenamefont {DeSutter}\ \emph {et~al.}(2019)\citenamefont
		{DeSutter}, \citenamefont {Tang},\ and\ \citenamefont
		{Francoeur}}]{DeSutter2019}%
	\BibitemOpen
	\bibfield  {author} {\bibinfo {author} {\bibfnamefont {J.}~\bibnamefont
			{DeSutter}}, \bibinfo {author} {\bibfnamefont {L.}~\bibnamefont {Tang}},\
		and\ \bibinfo {author} {\bibfnamefont {M.}~\bibnamefont {Francoeur}},\
	}\bibfield  {title} {\bibinfo {title} {A near-field radiative heat transfer
			device},\ }\href {https://doi.org/10.1038/s41565-019-0475-6} {\bibfield
		{journal} {\bibinfo  {journal} {Nat. Nanotechnol.}\ }\textbf {\bibinfo
			{volume} {14}},\ \bibinfo {pages} {751} (\bibinfo {year} {2019})}\BibitemShut
	{NoStop}%
	\bibitem [{\citenamefont {Shi}\ \emph {et~al.}(2019)\citenamefont {Shi},
		\citenamefont {Sun}, \citenamefont {Chen}, \citenamefont {He}, \citenamefont
		{Bao}, \citenamefont {Evans},\ and\ \citenamefont {He}}]{Shi2019}%
	\BibitemOpen
	\bibfield  {author} {\bibinfo {author} {\bibfnamefont {K.}~\bibnamefont
			{Shi}}, \bibinfo {author} {\bibfnamefont {Y.}~\bibnamefont {Sun}}, \bibinfo
		{author} {\bibfnamefont {Z.}~\bibnamefont {Chen}}, \bibinfo {author}
		{\bibfnamefont {N.}~\bibnamefont {He}}, \bibinfo {author} {\bibfnamefont
			{F.}~\bibnamefont {Bao}}, \bibinfo {author} {\bibfnamefont {J.}~\bibnamefont
			{Evans}},\ and\ \bibinfo {author} {\bibfnamefont {S.}~\bibnamefont {He}},\
	}\bibfield  {title} {\bibinfo {title} {Colossal enhancement of near-field
			thermal radiation across hundreds of nanometers between millimeter-scale
			plates through surface plasmon and phonon polaritons coupling},\ }\href
	{https://doi.org/10.1021/acs.nanolett.9b03269} {\bibfield  {journal}
		{\bibinfo  {journal} {Nano Lett.}\ }\textbf {\bibinfo {volume} {19}},\
		\bibinfo {pages} {8082} (\bibinfo {year} {2019})}\BibitemShut {NoStop}%
	\bibitem [{\citenamefont {Shi}\ \emph {et~al.}(2021)\citenamefont {Shi},
		\citenamefont {Chen}, \citenamefont {Xu}, \citenamefont {Evans},\ and\
		\citenamefont {He}}]{Shi2021}%
	\BibitemOpen
	\bibfield  {author} {\bibinfo {author} {\bibfnamefont {K.}~\bibnamefont
			{Shi}}, \bibinfo {author} {\bibfnamefont {Z.}~\bibnamefont {Chen}}, \bibinfo
		{author} {\bibfnamefont {X.}~\bibnamefont {Xu}}, \bibinfo {author}
		{\bibfnamefont {J.}~\bibnamefont {Evans}},\ and\ \bibinfo {author}
		{\bibfnamefont {S.}~\bibnamefont {He}},\ }\bibfield  {title} {\bibinfo
		{title} {Optimized colossal near-field thermal radiation enabled by
			manipulating coupled plasmon polariton geometry},\ }\href
	{https://doi.org/https://doi.org/10.1002/adma.202106097} {\bibfield
		{journal} {\bibinfo  {journal} {Adv. Mater.}\ }\textbf {\bibinfo {volume}
			{33}},\ \bibinfo {pages} {2106097} (\bibinfo {year} {2021})}\BibitemShut
	{NoStop}%
	\bibitem [{\citenamefont {Zhang}\ \emph {et~al.}(2024)\citenamefont {Zhang},
		\citenamefont {Wang}, \citenamefont {Jin}, \citenamefont {Zhou},
		\citenamefont {Gong},\ and\ \citenamefont {Zhao}}]{Wenbin2024}%
	\BibitemOpen
	\bibfield  {author} {\bibinfo {author} {\bibfnamefont {W.}~\bibnamefont
			{Zhang}}, \bibinfo {author} {\bibfnamefont {B.}~\bibnamefont {Wang}},
		\bibinfo {author} {\bibfnamefont {S.}~\bibnamefont {Jin}}, \bibinfo {author}
		{\bibfnamefont {J.}~\bibnamefont {Zhou}}, \bibinfo {author} {\bibfnamefont
			{Z.}~\bibnamefont {Gong}},\ and\ \bibinfo {author} {\bibfnamefont
			{C.}~\bibnamefont {Zhao}},\ }\bibfield  {title} {\bibinfo {title} {Colossal
			near-field radiative heat transfer mediated by coupled polaritons with an
			ultrahigh dynamic range},\ }\href
	{https://doi.org/https://doi.org/10.1002/adma.202405885} {\bibfield
		{journal} {\bibinfo  {journal} {Adv. Mater.}\ }\textbf {\bibinfo {volume}
			{36}},\ \bibinfo {pages} {2405885} (\bibinfo {year} {2024})}\BibitemShut
	{NoStop}%
	\bibitem [{\citenamefont {Luo}\ \emph {et~al.}(2024)\citenamefont {Luo},
		\citenamefont {Salihoglu}, \citenamefont {Wang}, \citenamefont {Li},
		\citenamefont {Kim}, \citenamefont {Liu}, \citenamefont {Li}, \citenamefont
		{Yu}, \citenamefont {Du},\ and\ \citenamefont {Shen}}]{Luo2024}%
	\BibitemOpen
	\bibfield  {author} {\bibinfo {author} {\bibfnamefont {X.}~\bibnamefont
			{Luo}}, \bibinfo {author} {\bibfnamefont {H.}~\bibnamefont {Salihoglu}},
		\bibinfo {author} {\bibfnamefont {Z.}~\bibnamefont {Wang}}, \bibinfo {author}
		{\bibfnamefont {Z.}~\bibnamefont {Li}}, \bibinfo {author} {\bibfnamefont
			{H.}~\bibnamefont {Kim}}, \bibinfo {author} {\bibfnamefont {X.}~\bibnamefont
			{Liu}}, \bibinfo {author} {\bibfnamefont {J.}~\bibnamefont {Li}}, \bibinfo
		{author} {\bibfnamefont {B.}~\bibnamefont {Yu}}, \bibinfo {author}
		{\bibfnamefont {S.}~\bibnamefont {Du}},\ and\ \bibinfo {author}
		{\bibfnamefont {S.}~\bibnamefont {Shen}},\ }\bibfield  {title} {\bibinfo
		{title} {Observation of near-field thermal radiation between coplanar
			nanodevices with subwavelength dimensions},\ }\href
	{https://doi.org/10.1021/acs.nanolett.3c04123} {\bibfield  {journal}
		{\bibinfo  {journal} {Nano Lett.}\ }\textbf {\bibinfo {volume} {24}},\
		\bibinfo {pages} {1502} (\bibinfo {year} {2024})}\BibitemShut {NoStop}%
	\bibitem [{\citenamefont {Tang}\ \emph {et~al.}(2024)\citenamefont {Tang},
		\citenamefont {Corr{\^{e}}a}, \citenamefont {Francoeur},\ and\ \citenamefont
		{Dames}}]{Tang2024}%
	\BibitemOpen
	\bibfield  {author} {\bibinfo {author} {\bibfnamefont {L.}~\bibnamefont
			{Tang}}, \bibinfo {author} {\bibfnamefont {L.~M.}\ \bibnamefont
			{Corr{\^{e}}a}}, \bibinfo {author} {\bibfnamefont {M.}~\bibnamefont
			{Francoeur}},\ and\ \bibinfo {author} {\bibfnamefont {C.}~\bibnamefont
			{Dames}},\ }\bibfield  {title} {\bibinfo {title} {Corner- and edge-mode
			enhancement of near-field radiative heat transfer},\ }\href
	{https://doi.org/10.1038/s41586-024-07279-2} {\bibfield  {journal} {\bibinfo
			{journal} {Nature}\ }\textbf {\bibinfo {volume} {629}},\ \bibinfo {pages}
		{67} (\bibinfo {year} {2024})}\BibitemShut {NoStop}%
	\bibitem [{\citenamefont {Zeng}\ \emph {et~al.}(2025)\citenamefont {Zeng},
		\citenamefont {Chen}, \citenamefont {Wu},\ and\ \citenamefont
		{Fu}}]{Zeng2025}%
	\BibitemOpen
	\bibfield  {author} {\bibinfo {author} {\bibfnamefont {C.}~\bibnamefont
			{Zeng}}, \bibinfo {author} {\bibfnamefont {S.}~\bibnamefont {Chen}}, \bibinfo
		{author} {\bibfnamefont {X.}~\bibnamefont {Wu}},\ and\ \bibinfo {author}
		{\bibfnamefont {C.}~\bibnamefont {Fu}},\ }\bibfield  {title} {\bibinfo
		{title} {Mode transition of near-field radiative heat transfer between two
			subwavelength cylinders},\ }\href
	{https://doi.org/10.1016/j.icheatmasstransfer.2025.109863} {\bibfield
		{journal} {\bibinfo  {journal} {Int. Commun. Heat Mass Transfer}\ }\textbf
		{\bibinfo {volume} {169}},\ \bibinfo {pages} {109863} (\bibinfo {year}
		{2025})}\BibitemShut {NoStop}%
	\bibitem [{\citenamefont {Corr{\^{e}}a}\ \emph {et~al.}(2026)\citenamefont
		{Corr{\^{e}}a}, \citenamefont {Tang},\ and\ \citenamefont
		{Francoeur}}]{Correa2026}%
	\BibitemOpen
	\bibfield  {author} {\bibinfo {author} {\bibfnamefont {M.~L.}\ \bibnamefont
			{Corr{\^{e}}a}}, \bibinfo {author} {\bibfnamefont {L.}~\bibnamefont {Tang}},\
		and\ \bibinfo {author} {\bibfnamefont {M.}~\bibnamefont {Francoeur}},\
	}\bibfield  {title} {\bibinfo {title} {Near-field radiative heat transfer in
			the dual nanoscale regime between polaritonic membranes},\ }\href
	{https://doi.org/10.1103/j7cj-4tl9} {\bibfield  {journal} {\bibinfo
			{journal} {Phys. Rev. Lett.}\ ,\ \bibinfo {pages} {01111}} (\bibinfo {year}
		{2026})}\BibitemShut {NoStop}%
	\bibitem [{\citenamefont {Volz}\ and\ \citenamefont
		{Ordonez-Miranda}(2024)}]{Volz24}%
	\BibitemOpen
	\bibfield  {author} {\bibinfo {author} {\bibfnamefont {S.}~\bibnamefont
			{Volz}}\ and\ \bibinfo {author} {\bibfnamefont {J.}~\bibnamefont
			{Ordonez-Miranda}},\ }\href@noop {} {\emph {\bibinfo {title} {Heat Transport
				Driven by Surface Electromagnetic Waves}}}\ (\bibinfo  {publisher} {Springer
		Cham},\ \bibinfo {address} {Switzerland},\ \bibinfo {year}
	{2024})\BibitemShut {NoStop}%
	\bibitem [{\citenamefont {Ordonez-Miranda}\ \emph {et~al.}(2026)\citenamefont
		{Ordonez-Miranda}, \citenamefont {Anufriev}, \citenamefont {Liñán-Abanto},
		\citenamefont {Coral}, \citenamefont {Nomura},\ and\ \citenamefont
		{Volz}}]{Ordonez2026}%
	\BibitemOpen
	\bibfield  {author} {\bibinfo {author} {\bibfnamefont {J.}~\bibnamefont
			{Ordonez-Miranda}}, \bibinfo {author} {\bibfnamefont {R.}~\bibnamefont
			{Anufriev}}, \bibinfo {author} {\bibfnamefont {R.~N.}\ \bibnamefont
			{Liñán-Abanto}}, \bibinfo {author} {\bibfnamefont {M.}~\bibnamefont
			{Coral}}, \bibinfo {author} {\bibfnamefont {M.}~\bibnamefont {Nomura}},\ and\
		\bibinfo {author} {\bibfnamefont {S.}~\bibnamefont {Volz}},\ }\bibfield
	{title} {\bibinfo {title} {Near-field thermal radiation between deep
			subwavelength membranes driven by corner and edge modes},\ }\href
	{https://doi.org/10.1063/5.0311645} {\bibfield  {journal} {\bibinfo
			{journal} {J. Appl. Phys.}\ }\textbf {\bibinfo {volume} {139}},\ \bibinfo
		{pages} {085108} (\bibinfo {year} {2026})}\BibitemShut {NoStop}%
	\bibitem [{\citenamefont {Anufriev}\ \emph {et~al.}(2022)\citenamefont
		{Anufriev}, \citenamefont {Wu}, \citenamefont {Ordonez-Miranda},\ and\
		\citenamefont {Nomura}}]{Anufriev2022}%
	\BibitemOpen
	\bibfield  {author} {\bibinfo {author} {\bibfnamefont {R.}~\bibnamefont
			{Anufriev}}, \bibinfo {author} {\bibfnamefont {Y.}~\bibnamefont {Wu}},
		\bibinfo {author} {\bibfnamefont {J.}~\bibnamefont {Ordonez-Miranda}},\ and\
		\bibinfo {author} {\bibfnamefont {M.}~\bibnamefont {Nomura}},\ }\bibfield
	{title} {\bibinfo {title} {Nanoscale limit of the thermal conductivity in
			crystalline silicon carbide membranes, nanowires, and phononic crystals},\
	}\href {https://doi.org/10.1038/s41427-022-00382-8} {\bibfield  {journal}
		{\bibinfo  {journal} {NPG Asia Mater.}\ }\textbf {\bibinfo {volume} {14}},\
		\bibinfo {pages} {35} (\bibinfo {year} {2022})}\BibitemShut {NoStop}%
	\bibitem [{\citenamefont {Chauvin}\ \emph {et~al.}(2012)\citenamefont
		{Chauvin}, \citenamefont {Hadj~Alouane}, \citenamefont {Anufriev},
		\citenamefont {Khmissi}, \citenamefont {Naji}, \citenamefont {Patriarche},
		\citenamefont {Bru-Chevallier},\ and\ \citenamefont {Gendry}}]{Chauvin2012}%
	\BibitemOpen
	\bibfield  {author} {\bibinfo {author} {\bibfnamefont {N.}~\bibnamefont
			{Chauvin}}, \bibinfo {author} {\bibfnamefont {M.~H.}\ \bibnamefont
			{Hadj~Alouane}}, \bibinfo {author} {\bibfnamefont {R.}~\bibnamefont
			{Anufriev}}, \bibinfo {author} {\bibfnamefont {H.}~\bibnamefont {Khmissi}},
		\bibinfo {author} {\bibfnamefont {K.}~\bibnamefont {Naji}}, \bibinfo {author}
		{\bibfnamefont {G.}~\bibnamefont {Patriarche}}, \bibinfo {author}
		{\bibfnamefont {C.}~\bibnamefont {Bru-Chevallier}},\ and\ \bibinfo {author}
		{\bibfnamefont {M.}~\bibnamefont {Gendry}},\ }\bibfield  {title} {\bibinfo
		{title} {Growth temperature dependence of exciton lifetime in wurtzite inp
			nanowires grown on silicon substrates},\ }\href
	{https://doi.org/10.1063/1.3674985} {\bibfield  {journal} {\bibinfo
			{journal} {Appl. Phys. Lett.}\ }\textbf {\bibinfo {volume} {100}},\ \bibinfo
		{pages} {011906} (\bibinfo {year} {2012})}\BibitemShut {NoStop}%
	\bibitem [{\citenamefont {Basu}\ \emph {et~al.}(2009)\citenamefont {Basu},
		\citenamefont {Zhang},\ and\ \citenamefont {Fu}}]{Basu09}%
	\BibitemOpen
	\bibfield  {author} {\bibinfo {author} {\bibfnamefont {S.}~\bibnamefont
			{Basu}}, \bibinfo {author} {\bibfnamefont {Z.~M.}\ \bibnamefont {Zhang}},\
		and\ \bibinfo {author} {\bibfnamefont {C.~J.}\ \bibnamefont {Fu}},\
	}\bibfield  {title} {\bibinfo {title} {Review of near-field thermal radiation
			and its application to energy conversion},\ }\href@noop {} {\bibfield
		{journal} {\bibinfo  {journal} {Int. J. Energy Res.}\ }\textbf {\bibinfo
			{volume} {33}},\ \bibinfo {pages} {1203} (\bibinfo {year}
		{2009})}\BibitemShut {NoStop}%
	\bibitem [{\citenamefont {Ordonez-Miranda}\ \emph {et~al.}(2024)\citenamefont
		{Ordonez-Miranda}, \citenamefont {Anufriev}, \citenamefont {Nomura},\ and\
		\citenamefont {Volz}}]{Ordonez24}%
	\BibitemOpen
	\bibfield  {author} {\bibinfo {author} {\bibfnamefont {J.}~\bibnamefont
			{Ordonez-Miranda}}, \bibinfo {author} {\bibfnamefont {R.}~\bibnamefont
			{Anufriev}}, \bibinfo {author} {\bibfnamefont {M.}~\bibnamefont {Nomura}},\
		and\ \bibinfo {author} {\bibfnamefont {S.}~\bibnamefont {Volz}},\ }\bibfield
	{title} {\bibinfo {title} {Dimensional crossover in thermal radiation: From
			three- to two-dimensional heat transfer between metallic membranes},\
	}\href@noop {} {\bibfield  {journal} {\bibinfo  {journal} {Phys. Rev. Appl.}\
		}\textbf {\bibinfo {volume} {22}},\ \bibinfo {pages} {L031006} (\bibinfo
		{year} {2024})}\BibitemShut {NoStop}%
	\bibitem [{\citenamefont {Biehs}\ \emph {et~al.}(2021)\citenamefont {Biehs},
		\citenamefont {Messina}, \citenamefont {Venkataram}, \citenamefont
		{Rodriguez}, \citenamefont {Cuevas},\ and\ \citenamefont
		{Ben-Abdallah}}]{Biehs21}%
	\BibitemOpen
	\bibfield  {author} {\bibinfo {author} {\bibfnamefont {S.-A.}\ \bibnamefont
			{Biehs}}, \bibinfo {author} {\bibfnamefont {R.}~\bibnamefont {Messina}},
		\bibinfo {author} {\bibfnamefont {P.~S.}\ \bibnamefont {Venkataram}},
		\bibinfo {author} {\bibfnamefont {A.~W.}\ \bibnamefont {Rodriguez}}, \bibinfo
		{author} {\bibfnamefont {J.~C.}\ \bibnamefont {Cuevas}},\ and\ \bibinfo
		{author} {\bibfnamefont {P.}~\bibnamefont {Ben-Abdallah}},\ }\bibfield
	{title} {\bibinfo {title} {Near-field radiative heat transfer in many-body
			systems},\ }\href@noop {} {\bibfield  {journal} {\bibinfo  {journal} {Rev.
				Mod. Phys.}\ }\textbf {\bibinfo {volume} {93}},\ \bibinfo {pages} {025009}
		(\bibinfo {year} {2021})}\BibitemShut {NoStop}%
	\bibitem [{\citenamefont {Rodriguez}\ \emph {et~al.}(2013)\citenamefont
		{Rodriguez}, \citenamefont {Reid},\ and\ \citenamefont
		{Johnson}}]{Rodriguez13}%
	\BibitemOpen
	\bibfield  {author} {\bibinfo {author} {\bibfnamefont {A.~W.}\ \bibnamefont
			{Rodriguez}}, \bibinfo {author} {\bibfnamefont {M.~T.~H.}\ \bibnamefont
			{Reid}},\ and\ \bibinfo {author} {\bibfnamefont {S.~G.}\ \bibnamefont
			{Johnson}},\ }\bibfield  {title} {\bibinfo {title}
		{Fluctuating-surface-current formulation of radiative heat transfer: Theory
			and applications},\ }\href@noop {} {\bibfield  {journal} {\bibinfo  {journal}
			{Phys. Rev. B}\ }\textbf {\bibinfo {volume} {88}},\ \bibinfo {pages} {054305}
		(\bibinfo {year} {2013})}\BibitemShut {NoStop}%
	\bibitem [{\citenamefont {Thompson}\ \emph {et~al.}(2018)\citenamefont
		{Thompson}, \citenamefont {Zhu}, \citenamefont {Mittapally}, \citenamefont
		{Sadat}, \citenamefont {Xing}, \citenamefont {McArdle}, \citenamefont
		{Qazilbash}, \citenamefont {Reddy},\ and\ \citenamefont
		{Meyhofer}}]{Thompson18}%
	\BibitemOpen
	\bibfield  {author} {\bibinfo {author} {\bibfnamefont {D.}~\bibnamefont
			{Thompson}}, \bibinfo {author} {\bibfnamefont {L.}~\bibnamefont {Zhu}},
		\bibinfo {author} {\bibfnamefont {R.}~\bibnamefont {Mittapally}}, \bibinfo
		{author} {\bibfnamefont {S.}~\bibnamefont {Sadat}}, \bibinfo {author}
		{\bibfnamefont {Z.}~\bibnamefont {Xing}}, \bibinfo {author} {\bibfnamefont
			{P.}~\bibnamefont {McArdle}}, \bibinfo {author} {\bibfnamefont {M.~M.}\
			\bibnamefont {Qazilbash}}, \bibinfo {author} {\bibfnamefont {P.}~\bibnamefont
			{Reddy}},\ and\ \bibinfo {author} {\bibfnamefont {E.}~\bibnamefont
			{Meyhofer}},\ }\bibfield  {title} {\bibinfo {title} {Hundred-fold enhancement
			in far-field radiative heat transfer over the blackbody limit},\ }\href@noop
	{} {\bibfield  {journal} {\bibinfo  {journal} {Nature}\ }\textbf {\bibinfo
			{volume} {561}},\ \bibinfo {pages} {216} (\bibinfo {year}
		{2018})}\BibitemShut {NoStop}%
	\bibitem [{\citenamefont {Tachikawa}\ \emph {et~al.}(2024)\citenamefont
		{Tachikawa}, \citenamefont {Ordonez-Miranda}, \citenamefont {Jalabert},
		\citenamefont {Wu}, \citenamefont {Anufriev}, \citenamefont {Guo},
		\citenamefont {Kim}, \citenamefont {Fujita}, \citenamefont {Volz},\ and\
		\citenamefont {Nomura}}]{Tachikawa24}%
	\BibitemOpen
	\bibfield  {author} {\bibinfo {author} {\bibfnamefont {S.}~\bibnamefont
			{Tachikawa}}, \bibinfo {author} {\bibfnamefont {J.}~\bibnamefont
			{Ordonez-Miranda}}, \bibinfo {author} {\bibfnamefont {L.}~\bibnamefont
			{Jalabert}}, \bibinfo {author} {\bibfnamefont {Y.}~\bibnamefont {Wu}},
		\bibinfo {author} {\bibfnamefont {R.}~\bibnamefont {Anufriev}}, \bibinfo
		{author} {\bibfnamefont {Y.}~\bibnamefont {Guo}}, \bibinfo {author}
		{\bibfnamefont {B.}~\bibnamefont {Kim}}, \bibinfo {author} {\bibfnamefont
			{H.}~\bibnamefont {Fujita}}, \bibinfo {author} {\bibfnamefont
			{S.}~\bibnamefont {Volz}},\ and\ \bibinfo {author} {\bibfnamefont
			{M.}~\bibnamefont {Nomura}},\ }\bibfield  {title} {\bibinfo {title} {Enhanced
			far-field thermal radiation through a polaritonic waveguide},\ }\href@noop {}
	{\bibfield  {journal} {\bibinfo  {journal} {Phys. Rev. Lett.}\ }\textbf
		{\bibinfo {volume} {132}},\ \bibinfo {pages} {186904} (\bibinfo {year}
		{2024})}\BibitemShut {NoStop}%
	\bibitem [{\citenamefont {Ordonez-Miranda}\ \emph {et~al.}(2025)\citenamefont
		{Ordonez-Miranda}, \citenamefont {Nomura},\ and\ \citenamefont
		{Volz}}]{Ordonez25}%
	\BibitemOpen
	\bibfield  {author} {\bibinfo {author} {\bibfnamefont {J.}~\bibnamefont
			{Ordonez-Miranda}}, \bibinfo {author} {\bibfnamefont {M.}~\bibnamefont
			{Nomura}},\ and\ \bibinfo {author} {\bibfnamefont {S.}~\bibnamefont {Volz}},\
	}\bibfield  {title} {\bibinfo {title} {Focusing surface phonon-polaritons for
			tunable thermal radiation},\ }\href@noop {} {\bibfield  {journal} {\bibinfo
			{journal} {Discover Nano}\ }\textbf {\bibinfo {volume} {20}},\ \bibinfo
		{pages} {15} (\bibinfo {year} {2025})}\BibitemShut {NoStop}%
\end{thebibliography}
%

\end{document}